\documentclass[aps,prb,reprint,twocolumn,amsmath,amssymb,citeautoscript,longbibliography]{revtex4-2}

\usepackage{graphicx}% Include figure files
\usepackage{dcolumn}% Align table columns on decimal point
\usepackage{bm}% bold math
\usepackage{latexsym,epsfig}
\usepackage{graphicx}
\usepackage{verbatim}
\usepackage{comment}
\usepackage{amsmath}
\usepackage{amssymb}
\usepackage{physics}
\usepackage{stmaryrd}
\usepackage{color}
\usepackage{epstopdf}
\usepackage{grffile}
\usepackage{lipsum}
\usepackage{enumitem}
\usepackage{ulem}
\usepackage[dvipsnames]{xcolor}
\DeclareGraphicsExtensions{.eps}

\newcommand{\beq}{\begin{equation}}
\newcommand{\eeq}{\end{equation}}
\newcommand{\bea}{\begin{eqnarray}}
\newcommand{\eea}{\end{eqnarray}}
\newcommand{\ben}{\begin{eqnarray*}}
\newcommand{\een}{\end{eqnarray*}}
\newcommand{\bfig}{\begin{figure}}
\newcommand{\efig}{\end{figure}}

\usepackage{hyperref}
\hypersetup{
    colorlinks=true,      
    urlcolor=blue,
    citecolor=blue,
    linkcolor=blue
}

\usepackage{booktabs}
\usepackage{array}
\usepackage{lipsum} % For dummy text, optional
\usepackage{array} % Add this to your preamble

\begin{document}

\title{Emergence of charge and spin current in non-Hermitian quantum ring}

\author{Soumya Ranjan Padhi}
\email{padhisoumyaranjan778@gmail.com}
\affiliation{School of Physical Sciences, National Institute of Science Education and Research, Jatni 752050, India}
\affiliation{Homi Bhabha National Institute, Training School Complex, Anushaktinagar, Mumbai 400094, India}
\author{Souvik Roy}
\email{souvikroy138@gmail.com }
\affiliation{School of Physical Sciences, National Institute of Science Education and Research, Jatni 752050, India}
\affiliation{Homi Bhabha National Institute, Training School Complex, Anushaktinagar, Mumbai 400094, India}
\author{Tapan Mishra}
\email{mishratapan@gmail.com}
\affiliation{School of Physical Sciences, National Institute of Science Education and Research, Jatni 752050, India}
\affiliation{Homi Bhabha National Institute, Training School Complex, Anushaktinagar, Mumbai 400094, India}

\date{\today}

\begin{abstract}
We investigate the charge and spin transport in a non-Hermitian ring of electrons subject to an external Zeeman field. By introducing non-Hermiticity through anti-Hermitian hopping in the nearest neighbour bonds, we demonstrate that anti-Hermiticity, along with the applied Zeeman field significantly modify the energy spectrum and strongly influence transport properties. As a result, we obtain that when antiferromagnetic Zeeman field is considered, a finite charge current emerges in both the real and imaginary parts of the current, which are in contrast to the ferromagnetic case where only the imaginary current exist. On the other hand, in both cases, the spin current vanishes. Interestingly, we reveal an emergence and strong enhancement of spin currents under balanced spin population upon introducing quasiperiodicity in the presence of antiferromagnetic ordering. At the same time, the charge current also exhibits substantial enhancement due to quasiperiodic modulation. These results highlight non-Hermitian quantum rings as versatile platforms for unconventional spin-charge transport.
\end{abstract}

\maketitle

\section{Introduction}

Non-Hermitian (NH) quantum systems have emerged as fertile ground for uncovering novel physical phenomena that have no counterparts in conventional Hermitian settings~\cite{r1,r2,r3,r4,r5,r6,r7,r8,r9,r10,r11,r12,r13,r14,r15}. A major driving force behind this surge of interest is the realization that effective non-Hermiticity naturally arises when a quantum system interacts with its surrounding in the form of gain–loss processes, non-reciprocal dynamics, or complex on-site potentials~\cite{r16,r17,r18,r19,r20,r21,r22,r23,r24,r25,r26,r27,r28,r29,r30,r31,r32}. Remarkably, under appropriate symmetry constraints, most notably parity-time (PT) symmetry, such systems can sustain entirely real energy spectra despite being intrinsically non-Hermitian~\cite{r1, r2}. This counterintuitive possibility has propelled NH and PT-symmetric systems to the forefront of contemporary research across a wide range of physical platforms.

Beyond spectral properties, transport phenomena in NH systems have attracted increasing attention~\cite{nhpc4, nhpc3, nhpc5, nhpc2, nhpc1}. In open quantum systems, the presence of gain and loss profoundly alters localization behavior, transmission characteristics, and current-carrying capabilities. Exceptional points, in particular, have been shown to dramatically amplify transport responses and to induce strong circulating currents, which in turn generate effective magnetic fields. Such effects are especially intriguing in loop geometries, where quantum interference plays a decisive role. While circular currents in mesoscopic rings threaded by an Aharonov-Bohm (AB) flux~\cite{pc1,pc2} have long been studied within Hermitian and disordered frameworks, their behavior in non-Hermitian environments, especially under structured gain–loss patterns, remains far from fully understood.

According to the seminal work of B\"uttiker and collaborators, a current induced in a ring by an AB flux can persist even after the flux is removed, highlighting the fundamentally equilibrium nature of persistent currents\cite{pc3,pc4,pc5,pc6,pc7,pc8,pc9,pc10,pc11}. In Hermitian disordered systems, these currents are typically suppressed as disorder strength increases~\cite{dis00,dis01,dis02}. However, it has also been demonstrated that specific correlations between site energies and hopping dimerization can counteract localization and even enhance the current. Despite these advances, the role of non-Hermitian effects, particularly the interplay between environmental gain-loss mechanisms and spin ordering, has received comparatively little attention. An especially intriguing aspect is the possibility that a system, although non-Hermitian by construction, may support a purely real current up to a critical value of the non-Hermiticity parameter, beyond which complex contributions emerge. This opens a fundamentally new route for controlling transport using environmental coupling.

Guided by these motivations, we explore the magnetic transport in a ring lattice with anti-Hermitian hopping in the presence of Zeeman field and then study the role of onsite disorder in the form of quasiperiodic modulation on the transport properties. The specific choice of anti-Hermitian hopping introduces non-Hermiticity in the system and most importantly an effective flux that threads through the ring which plays a crucial role in generating transport in the system when coupled to the Zeeman field.
%In these models, non-Hermiticity arises from asymmetric hopping amplitudes, while quasiperiodic or staggered modulations introduce additional competing length and energy scales. 
We show that in the absence of disorder, the real component of the current vanishes, while a finite imaginary current persists in both the nonmagnetic and ferromagnetic cases. In contrast, the antiferromagnetic configuration supports finite real currents in both spin channels, giving rise to a nonzero charge current. Since the up- and down-spin contributions are identical, the net spin current vanishes. Upon introducing osnite quasiperiodic modulation, the symmetry between the spin-resolved currents is lifted, leading to the simultaneous emergence of a finite spin current alongside the charge current. This framework allows us to systematically isolate and contrast the individual and combined effects of   non-Hermitian hopping, Zeeman coupling and quasiperiodicity on the spectral characteristics and transport response of closed-loop systems.

The remainder of the paper is organized as follows. In Sec.~II, we introduce the model Hamiltonian and describe the formalism employed to compute the energy spectra and transport currents. Section~III is devoted to a detailed analysis of the numerical results and their physical implications. Finally, Sec.~IV summarizes the main conclusions and outlines possible directions for future research.

% \textcolor{red}{We have to arrange this properly}
% The remainder of the paper is organized as follows. In Sec. II, we introduce the Hamiltonian and outline the formalism used to compute energy spectra and currents. Section III presents a detailed discussion of the numerical results and their physical implications. Finally, Sec. IV summarizes the main findings and highlights possible directions for future research.
\begin{figure}[t]
    \centering    \includegraphics[width=0.85\linewidth]{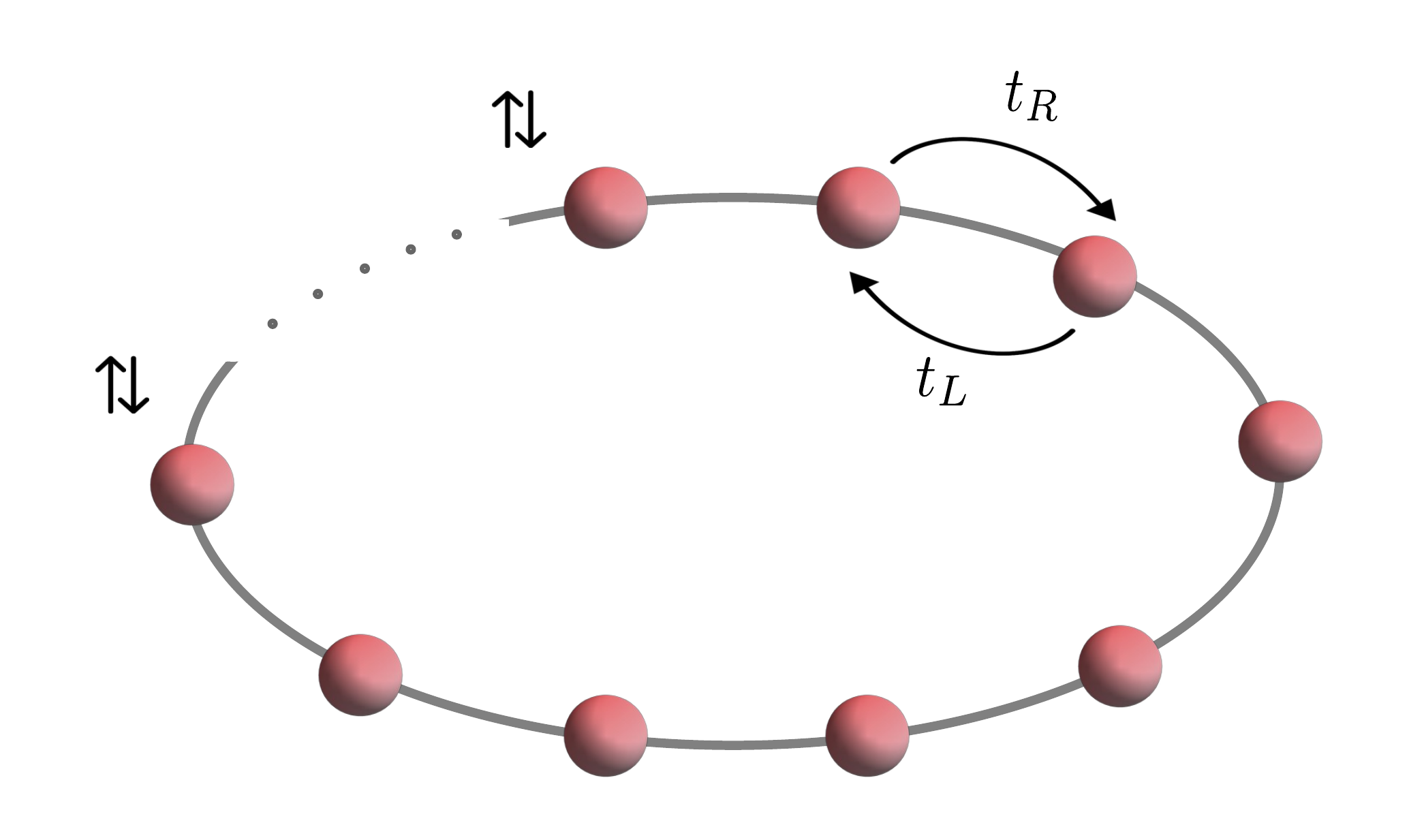}
    \caption{ Schematic illustration of an anti-Hermitian spinful quantum ring. The arrows attached to individual lattice sites represent the local spins.  
    %Directional asymmetry in electron motion is encoded through anti-Hermitian hopping amplitudes, with
    $t_R$ and $t_L$ denotes the directional hopping processes between the nearest neighbour sites.
}
    \label{fig:fig1}
\end{figure}
\section{Model and approach}
\label{sec:model}
In this section, we introduce the schematic structure of the model and present the corresponding tight-binding Hamiltonian. We consider a system of spinful fermions on a ring lattice possessing directional hopping amplitudes, as shown in Fig.~\ref{fig:fig1}. This system when subjected to a Zeeman field can be expressed by the Hamiltonian given as 
\begin{align}
 \mathcal{H} = & \sum_{j,\sigma \in \{\uparrow,\downarrow\}}
\left(
t_L\, c_{j\sigma}^\dagger c_{j+1,\sigma}
+ t_R\, c_{j+1,\sigma}^\dagger c_{j\sigma}
\right) \nonumber \\ &
+ \sum_j
\left(\epsilon_j + (-1)^j h_z \right)
c_{j\uparrow}^\dagger c_{j\uparrow}
\nonumber \\ 
& + \sum_j\left(\epsilon_j - (-1)^j h_z \right)
c_{j\downarrow}^\dagger c_{j\downarrow}.
\end{align}
where $c_{j\sigma}^\dagger$ ($c_{j\sigma}$) creates (annihilates) a fermion with spin $\sigma\in \{\uparrow,\downarrow\}$ at site $j$. The hopping amplitudes in the clockwise and anticlockwise directions are denoted by $t_R$ and $t_L$, respectively. The parameter \( h_z \) denotes the strength of the Zeeman field, which couples to the spin degree of freedom. For \( j = 0 \), the Zeeman field is spatially uniform across the lattice, corresponding to a ferromagnetic (FM) configuration. For \( j = 1, \ldots, L \), the Zeeman field alternates in sign between odd and even lattice sites, realizing an antiferromagnetic (AFM) texture.  $\epsilon_j$ represents an onsite quasiperiodic modulation of the form 
$\epsilon_j = \lambda \cos(2\pi \beta j + \delta)$ with $\beta = (\sqrt{5}-1)/2$ ensuring quasiperiodicity; throughout this work we set $\delta = 0$.

We choose anti-Hermitian hopping amplitudes by imposing the condition $t_L=-t_R^{*}$ which makes the system non-Hermitian. Following Ref.~\cite{nhpc4}, the hopping amplitudes can be parametrized as $t_L=t_0+i\eta$ and $t_R=-t_0+i\eta$, where $t_0$ and $\eta$ belongs to the real number. Expressing these amplitudes in polar form as $t_{L} = \sqrt{t_{0}^{2}+\eta^{2}}\,e^{i\phi}$ and $t_{R} = -\sqrt{t_{0}^{2}+\eta^{2}}\,e^{-i\phi}$, where $\phi=\tan^{-1}(\eta/t_{0})$ acts as the phase corresponding to an effective gauge field. This phase accumulation can be interpreted as a synthetic magnetic flux threading the ring, which emerges intrinsically from the non-Hermitian hopping structure rather than from an external gauge field. Such an effective flux provides a natural route to control transport properties and circulating currents in the ring geometry. To emulate a uniform auxiliary flux in a ring geometry, this phase may be scaled with the system size as $\phi^{\prime} = \frac{2\pi}{L}\tan^{-1}(\eta/t_{0})=\frac{2\pi}{L}\phi$. Introducing $t=\sqrt{t_{0}^{2}+\eta^{2}}$, the hopping terms simplify to $t_{L}=t e^{i\phi^{\prime}}$ and $t_{R}=-t e^{-i\phi^{\prime}}$. 
%The specific functional forms of $\epsilon_j$ are introduced later, in the relevant subsections, where different types of disorder are discussed.
%and the anti-Hermitian nearest-neighbor hopping. 
%The hopping amplitudes in the clockwise and anticlockwise directions are denoted by $t_R$ and $t_L$, respectively.
%The complex nature of these hopping amplitudes gives rise to effective phase factors accumulated by charge carriers encircling the ring. 

In the following we will first discuss the interplay of non-Hermitian hopping, Zeeman field, and quantum interference induced by the effective flux in detail and then study the effect of disorder on the transport properties. 
%Within the tight-binding framework, the Hamiltonian is given by
% AAH potential, which is kept in a general form as follows, 
% \begin{equation}
%     \epsilon_j = \lambda cos(2\pi\beta j + \delta)
% \end{equation}
% where $\beta = (\sqrt{5} -1) / 2$ known as the inverse golden mean ratio, which ensures the quasiperiodicity in the system. Throughout this work, we set $\delta=0$ for simplicity.

\begin{figure*}[t]
    \centering
    \includegraphics[width=1.0\linewidth]{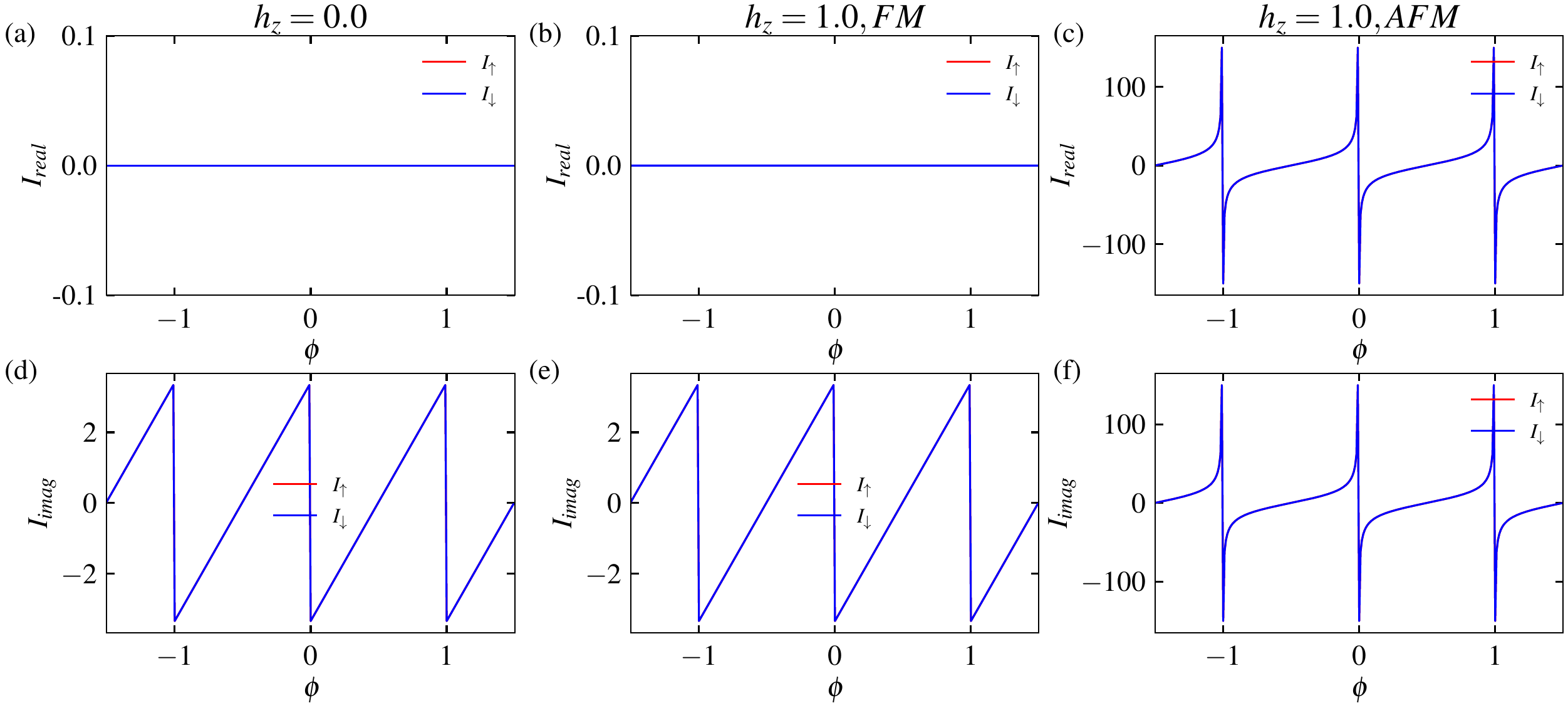}
    \caption{Real $(I_{\mathrm{real}})$ and imaginary $(I_{\mathrm{imag}})$ components of the current as functions of the flux $\phi$. Panels (a) and (d) correspond to the case without a Zeeman field. Panels (b) and (e) show the results in the presence of a Zeeman field $h_z = 1.0$ for a ferromagnetic configuration,
    while panels (c) and (f) correspond to $h_z = 1.0$ with an antiferromagnetic configuration.
    The solid blue and red curves represent the spin-up ($I_\uparrow$) and spin-down ($I_\downarrow$) currents, respectively.
}
    \label{fig:fig2}
\end{figure*}

To characterize transport in the non-Hermitian ring, we analyze the current, defined through the sensitivity of the ground-state energy to the applied magnetic flux~\cite{nhpc4},
\begin{equation}
I_{\sigma}^{r/i} = -c\,\frac{\partial E_{0,\sigma}^{r/i}}{\partial \phi},
\label{eq:current_def}
\end{equation}
where \(E_{0,\sigma}^{r/i}\) denotes the real or imaginary part of the spin-resolved ground-state energy for spin \(\sigma\in \{\uparrow,\downarrow\}\), and \(c\) is an appropriate proportionality constant.
Owing to the complex spectrum of the non-Hermitian Hamiltonian, the real and imaginary components of the spectrum define distinct Fermi surfaces, each constructed exclusively from the corresponding real or imaginary eigenvalues. Consequently, the ground-state energies are evaluated independently within the real- and imaginary-energy sectors, and are computed as
\begin{equation}
E_0^{r/i} = \sum_{m=1}^{N_e} E_m^{r/i},
\end{equation}
where \(E_m^{r/i}\) is the real or imaginary part of the \(m\)-th single-particle eigenvalue and \(N_e\) is the number of occupied states. The total charge and spin currents follow from the symmetric and antisymmetric combinations of the spin-resolved responses,
\begin{equation}
I_{\mathrm{c}}^{r/i} = \sum_{\sigma} I_{\sigma}^{r/i},
\qquad
I_{\mathrm{s}}^{r/i} = I_{\uparrow}^{r/i} - I_{\downarrow}^{r/i},
\label{eq:charge_spin_current}
\end{equation}
which respectively characterize collective charge transport and spin-selective current flow. 
%This compact framework enables a transparent analysis of how quasiperiodicity, non-Hermiticity, and spin-dependent effects jointly govern transport in the system.

\begin{figure*}[!htbp]
    \centering
    \includegraphics[width=1.0\linewidth]{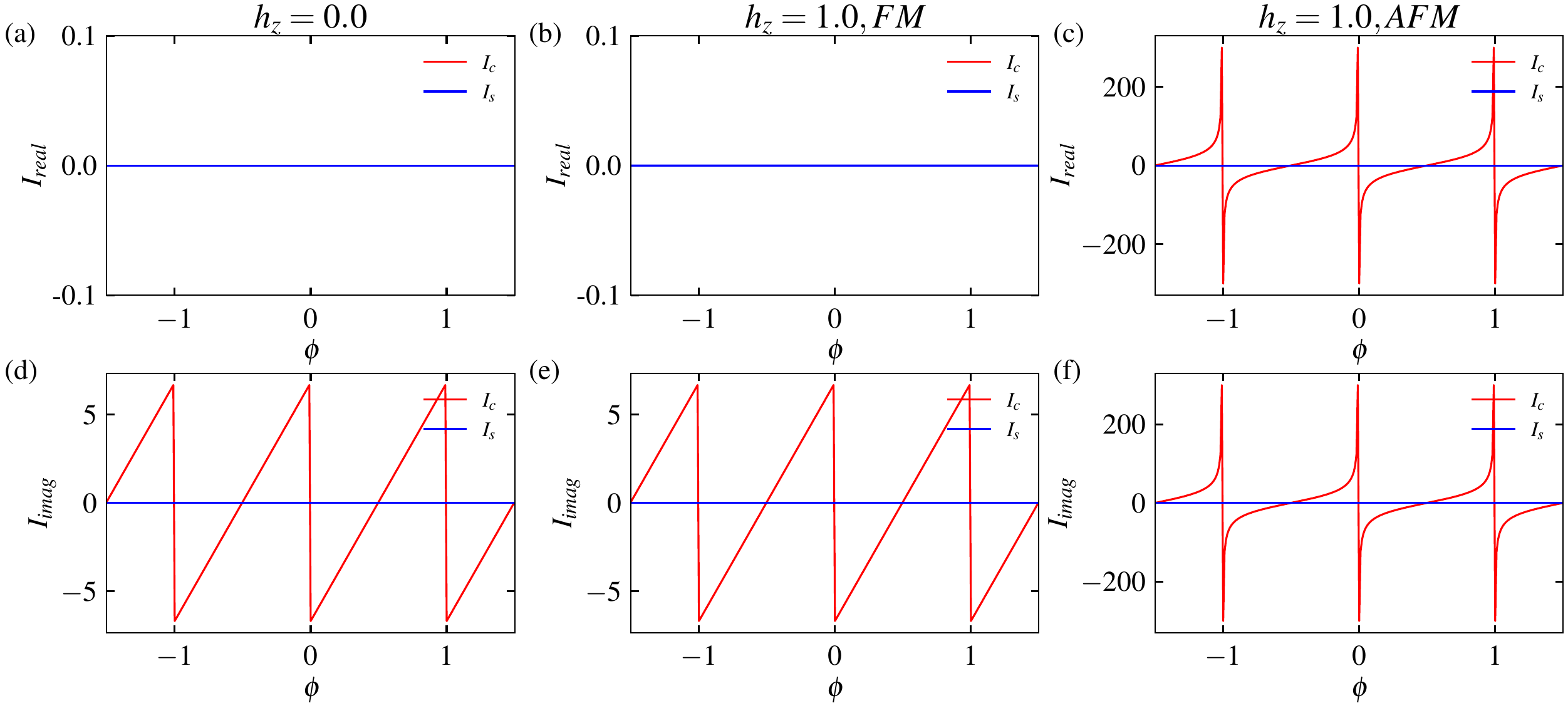}
    \caption{
Flux dependence of the real $(I_{\mathrm{real}})$ and imaginary $(I_{\mathrm{imag}})$ components of the current for different Zeeman-field configurations. Panels (a) and (d) show the response in the absence of a Zeeman field. Panels (b) and (e) correspond to a uniform (ferromagnetic) Zeeman field with strength $h_z = 1.0$, while panels (c) and (f) display the results for a staggered (antiferromagnetic) Zeeman field of the same strength. The charge ($I_c$) and spin ($I_s$) currents are indicated by the blue and red solid lines, respectively.
}

    \label{fig:fig3}
\end{figure*}

\section{Results}

In this section we present the main results of our study which is divided into two parts. We begin by considering a situation where both the Zeemen field ($h_z$) and onsite potential ($\epsilon_j$) are absent. This case allows us to clearly establish the intrinsic transport properties of the system only due to the anti-Hermitian hopping. We then introduce the Zeemen field and examine its effect on the current for both FM and AFM spin configurations. Finally, we incorporate the quasiperiodic disorder and investigate how this modulation modifies the transport behavior. Throughout our analysis, all calculations are performed for $144$ sites and the hopping amplitude is fixed to $t=1.0$. 
% Unless stated otherwise, we consider a system size of $L=144$ lattice sites for the standard AAH model. 
Transport properties are characterized by computing the absolute maximum current, obtained by scanning the auxiliary magnetic flux over the range $-2 \leq \phi \leq 2$. All calculations are performed at half filling unless explicitly specified. 
%Energies, including hopping and onsite terms, are measured in electron-volts (eV), and the resulting currents are expressed in units of $\mu\mathrm{A}$.

\subsection{Clean limit}
Here, we focus on the clean limit by setting $\epsilon_j = 0$ and examine the transport properties of the system both in the absence and in the presence of the Zeeman field $h_z$.

Figure~\ref{fig:fig2} displays the spin-resolved current as a function of $\phi$ for different Zeeman-field configurations. 
The real ($I_{real}$) and imaginary ($I_{imag}$) components of the current are shown in the upper and lower panels, respectively, reflecting the nature of the energy spectrum.
In the nonmagnetic ($h_z = 0$) and uniform FM cases, the real parts of the current vanishes identically for both the spin components as shown in Fig.~\ref{fig:fig2}(a) and Fig.~\ref{fig:fig2}(b), respectively. 
% This behavior is consistent with the purely imaginary character of the energy dispersion in these configurations. 
In contrast, upon introducing a staggered Zeeman field corresponding to AFM configuration, a finite real current emerges, as illustrated in Fig.~\ref{fig:fig2}(c), signaling a qualitative modification of the spectral properties.
The imaginary component of the current exhibits a sawtooth-like dependence on $\phi$. 
Although finite in the nonmagnetic and FM cases [Fig.~\ref{fig:fig2}(d) and Fig.~\ref{fig:fig2}(e)], its magnitude remains relatively small. 
Remarkably, in the AFM configuration, the imaginary current is significantly enhanced, as shown in Fig.~\ref{fig:fig2}(f), indicating a strong amplification of the transport response induced by the staggered Zeeman field.

To elucidate the origin of this behavior, we examine the band dispersion and the associated current by representing the Hamiltonian in momentum space through a Fourier transformation of the fermionic operators. We introduce the spinor in the combined sublattice and spin basis as
\begin{equation}
\Psi_k =
\big(
c_{kA\uparrow},
c_{kA\downarrow},
c_{kB\uparrow},
c_{kB\downarrow}
\big)^{T},
\label{eq:spinor}
\end{equation}
where $A$ and $B$ denote the two sublattices within a unit cell. The momentum-space operators are defined as
\begin{equation}
c_{k\alpha\sigma}
=
\frac{1}{\sqrt{N}}
\sum_{j}
e^{-ikj}
c_{j\alpha\sigma},
\qquad
\sigma = \uparrow, \downarrow,
\label{eq:fourier1}
\end{equation}
with $\alpha = A, B$ and $N$ being the total number of unit cells. For a ring geometry with periodic boundary conditions, the inverse Fourier transformation reads
\begin{equation}
c_{j\alpha\sigma}
=
\frac{1}{\sqrt{N}}
\sum_{k}
e^{ikj}
c_{k\alpha\sigma},
\quad
k = \frac{2\pi n}{N},
n = 0, 1, \dots, N-1.
\label{eq:fourier2}
\end{equation}
the Hamiltonian takes the Bloch form
\begin{equation}
H = \sum_k \Psi_k^\dagger \, \mathcal{H}(k) \, \Psi_k.
\end{equation}
\begin{equation}
\mathcal{H}(k)
=
\begin{pmatrix}
 h_z & 0 & f(k) & 0  \\
 0 & -h_z & 0 & f(k) \\
 g(k) & 0 & h_z & 0 \\
 0 & g(k) & 0 & -h_z  \\
\end{pmatrix},
\label{eq:FM_Bloch_fg}
\end{equation}
where the off-diagonal hopping functions are defined as
\begin{align}
f(k) &= t_L + t_R e^{-ik}, \\
g(k) &= t_R + t_L e^{ik}.
\end{align}
From this Bloch Hamiltonian, one can directly diagonalize $\mathcal{H}(k)$ to obtain the energy dispersion. 
The resulting eigenvalues in absence of the Zeeman field ($h_z=0$) are found to be purely imaginary and are given by,
\begin{equation}
E_{\pm}(k) = 
\pm 2 i t
\sin\!\left(\frac{k}{2}+\frac{2\pi\phi}{L}\right).
\end{equation}
From the above expression, since the energy has no real component, the real part of the current vanishes identically for both the spin components. The current arises solely from the imaginary part of the spectrum and is obtained as
\begin{equation}
I =
\sum_k I_{\sigma,\pm}(k) =
\sum_k
\pm \frac{4\pi}{L} c i t
\cos\!\left(\frac{k}{2}+\frac{2\pi\phi}{L}\right).
\end{equation}
This leads to a finite imaginary current with a characteristic sawtooth-like dependence on the magnetic flux, as shown in Fig.~\ref{fig:fig2}(a) and Fig.~\ref{fig:fig2}(d). The identical behavior of the spin-up and spin-down currents reflects the absence of spin splitting in this regime.

When a uniform Zeeman field corresponding to the FM configuration is turned on the dispersion relation is given as,
\begin{equation}
E_{s,\pm}(k)
=s h_z
\pm
2 i t
\sin\!\left(\frac{k}{2}+\frac{2\pi \phi}{L}\right),
\qquad s=\pm 1 .
\end{equation}
The Zeeman field lifts the spin degeneracy by shifting the two spin bands by 
$\pm h_z$. However, this shift is independent of the magnetic flux and therefore does not contribute to the current. Consequently, the expression for the current remains identical to the zero-field case.
As a result, the real component of the current remains zero, while the imaginary component retains its flux-dependent sawtooth structure, as shown in Fig.~\ref{fig:fig2}(b) and Fig.~\ref{fig:fig2}(e), respectively. This demonstrates that, in the FM case, the Zeeman field acts as a rigid energy shift and does not qualitatively modify the transport properties induced by the non-Hermitian hopping.

However, in the presence of a staggered Zeeman field, the structure of the Bloch Hamiltonian changes significantly and is given by,
\begin{equation}
\mathcal{H}(k)
=
\begin{pmatrix}
 h_z & 0 & f(k) & 0 \\
 0 & -h_z & 0 & f(k) \\
 g(k) & 0 & -h_z & 0 \\
 0 & g(k) & 0 & h_z
\end{pmatrix}.
\label{eq:AFM_Bloch_fg}
\end{equation}
The staggered nature of the Zeeman field couples different sublattice sectors and 
qualitatively modifies the band structure. Consequently, a markedly different transport behavior emerges in the AFM configuration, as illustrated in 
Fig.~\ref{fig:fig2}(c) and Fig.~\ref{fig:fig2}(f). Here, the Zeeman field competes with the anti-Hermitian hopping, giving rise to a dispersion relation that can be either real or purely imaginary, depending on momentum and flux. The dispersion relation is given by,
\begin{equation}
\begin{cases}
E_{s,\pm}(k) = \pm \sqrt{\Delta(k+\phi^{\prime})}, 
& |h_z|\ge 2|t|\left|\sin\!\left(\frac{k}{2}+\phi^{\prime}\right)\right|,\\
E_{s,\pm}(k) = \pm i \sqrt{-\Delta(k+\phi^{\prime})},
& |h_z|< 2|t|\left|\sin\!\left(\frac{k}{2}+\phi^{\prime}\right)\right|,
\end{cases}
\end{equation}
where, $\Delta(k+\phi^{\prime})=h_z^2 - 4t^2\sin^2(k/2 + \phi^\prime)$ and $\phi^{\prime}=2\pi \phi / L$.
Thus, despite the non-Hermitian hopping, the AFM configuration supports real energy bands whenever the staggered Zeeman field dominates. In this regime, the current is purely real and is given by
\begin{equation}
I = \sum_k I_{\pm}(k) = \sum_k 
\pm c
\frac{
\frac{4\pi}{L} t^2
\sin\!\left(k+\frac{4\pi\phi}{L}\right)
}{
\sqrt{
h_z^2-
4 t^2
\sin^2\!\left(\frac{k}{2}+\frac{2\pi\phi}{L}\right)
}
}.
\end{equation}
In contrast, when the anti-Hermitian hopping dominates, the spectrum becomes purely imaginary and the corresponding expression for current reads
\begin{equation}
I = \sum_k I_{\pm}(k) =
\sum_k \mp i c
\frac{
\frac{4\pi}{L} t^2
\sin\!\left(k+\frac{4\pi\phi}{L}\right)
}{
\sqrt{
4 t^2
\sin^2\!\left(\frac{k}{2}+\frac{2\pi\phi}{L}\right)-
h_z^2
}
}.
\end{equation}

Accordingly, both real and imaginary components of the current appear, depending on the flux and the relative strength of the staggered Zeeman field. Importantly, in the AFM configuration, the spin-up and spin-down channels contribute finite and distinct currents, reflecting the genuine spin-dependent transport induced by the interplay of non-Hermiticity and staggered magnetic field.

% Having established the behavior of the spin-resolved currents, we now turn to the charge and spin currents shown in Fig. 3. The charge current is defined as the sum of the spin-up and spin-down contributions, while the spin current is given by their difference. As seen from Fig. 3, a finite charge current appears only in the antiferromagnetic configuration of the Zeeman field. This behavior directly reflects the fact that, in the AFM case, the real and imaginary components of the spin-resolved currents become flux dependent and do not cancel in the charge channel. In contrast, the spin current remains identically zero for all flux values and Zeeman-field configurations, since the spin-up and spin-down currents are equal in magnitude. This confirms that, despite the presence of non-Hermiticity and staggered magnetic fields, the system does not support a net spin transport, while allowing for a finite charge response only in the AFM phase. So the question naturally arises how to emerge spin current in the system?
Having established the behavior of the spin-resolved currents, we now examine the corresponding charge and spin currents shown in Fig.~\ref{fig:fig3}. The charge current is defined as the sum of the spin-up and -down contributions, whereas the spin current is given by their difference. As evident from Fig.~\ref{fig:fig3}, a finite charge current emerges exclusively in the AFM configuration of the Zeeman field. This reflects the fact that, in the AFM case, the real and imaginary components of the spin-resolved currents acquire a flux dependence and no longer cancel in the charge sector. By contrast, the spin current remains identically zero for all flux values and magnetic configurations, since the spin-up and spin-down currents are equal in magnitude. Thus, despite the presence of non-Hermiticity and staggered magnetic fields, the system supports a finite charge response only in the AFM configuration while exhibiting no net spin transport. 

In what follows, we will show when the onsite quasiperiodic potential is turned on it imparts dramatic effect on the transport properties. 
%This naturally raises the question of how a finite spin current can be induced in the system. Which naturally motivates an exploration of the different onsite disorder potential to assess the possibility of spin transport.

\subsection{Effect of onsite quasiperiodic modulation}

\begin{figure}[t]
\centering   \includegraphics[width=1.0\linewidth]{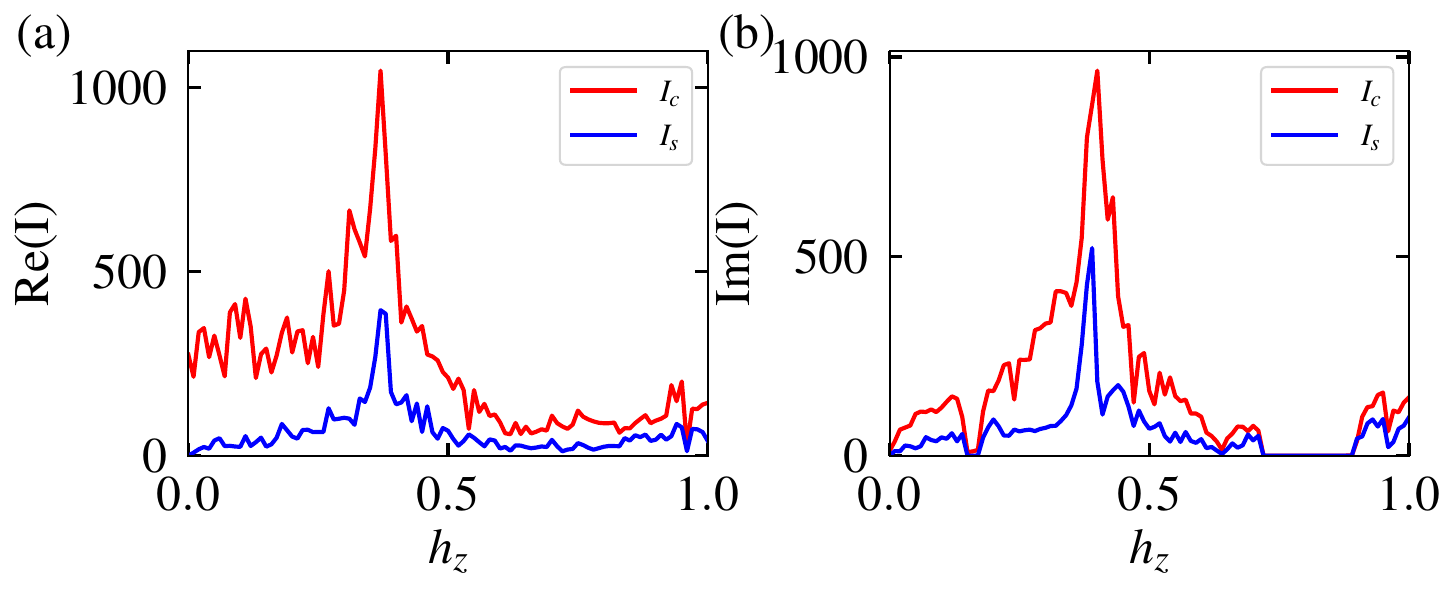}
\caption{(a) Real and (b) imaginary parts of the current as a function of the field strength $h_z$ for $\lambda=1$. The red and blue curves correspond to the charge and spin currents, $I_c$ and $I_s$, respectively. }
    \label{fig:aahhz}
\end{figure}

% We begin with the Aubry-Andr\'e-Harper (AAH) model to establish a clear baseline for the effects of quasiperiodicity in the non-Hermitian setting. 
As mentioned in the previous section, we assume Aubry-Andr\'e-Harper (AAH) type onsite quasiperiodic modulation. 
To elucidate the role of the Zeeman field, we examine the evolution of the charge and spin currents as functions of the Zeeman field strength \( h_z \), while keeping the disorder strength fixed at \( \lambda = 1 \). This allows us to isolate the influence of the magnetic field on transport and to identify characteristic signatures arising from the interplay between quasiperiodicity and non-Hermiticity. 
% \sout{A useful point of reference is the behavior in the absence of non-Hermiticity: for a ferromagnetic (FM) configuration, the spin current strictly vanishes, and this remains true even when the magnetic background is antiferromagnetic (AFM), provided the Hamiltonian is Hermitian.} 
It is well known that at half filling, or more generally when the populations of spin-up and spin-down electrons are equal, a Hermitian system without electron-electron interaction exhibits identical energy spectra for the two spin sectors, reflecting an exact spin degeneracy. Introducing interaction effects in this balanced filling regime does lead to a slight splitting between the eigenenergies of spin-up and spin-down electrons. However, this separation remains quantitatively small and consequently generates only a negligible spin current. Even when magnetic correlations such as FM or AFM ordering are incorporated within a Hermitian framework, the resulting modification of the spectrum is typically insufficient to robustly lift the spin degeneracy or to induce a pronounced asymmetry between the spin-resolved eigenvalues. As a result, purely Hermitian systems under these conditions fail to support sufficient spin currents, thereby underscoring the inherent limitations of Hermitian spin engineering in balanced fillings. Thus, to support sufficiently higher spin transport requires not only AFM order but also a non-Hermitian environment capable of breaking the symmetry between the two spin channels. Once non-Hermiticity is introduced together with moderate quasiperiodic modulation, the response changes qualitatively. In Fig.~\ref{fig:aahhz}, we show the real and imaginary components of the current, with the red and blue curves representing the charge and spin sectors, respectively. As shown in the plots, both $I_c$ and $I_s$ increase with $h_z$ in a distinctly nonlinear fashion up to roughly $h_z \approx 0.37$, reflecting the gradual field-induced imbalance between the spin-up and spin-down transport channels and the corresponding enhancement of the charge response. Beyond this intermediate field scale, the currents decrease as the Zeeman splitting becomes strong enough to suppress coherent hopping in one of the channels. Interestingly, for even stronger fields ($h_z \gtrsim 0.75$) both currents exhibit a weak re-emergence, signaling a second regime opening alternative pathways for transport. This re-entrant behavior is most clearly visible in the imaginary component of the current shown in Fig.~\ref{fig:aahhz}(b), underscoring the subtle and highly nonmonotonic manner in which magnetic field and quasiperiodicity together shape the non-Hermitian current landscape.

\begin{figure}[t]
    \centering   \includegraphics[width=1.0\linewidth]{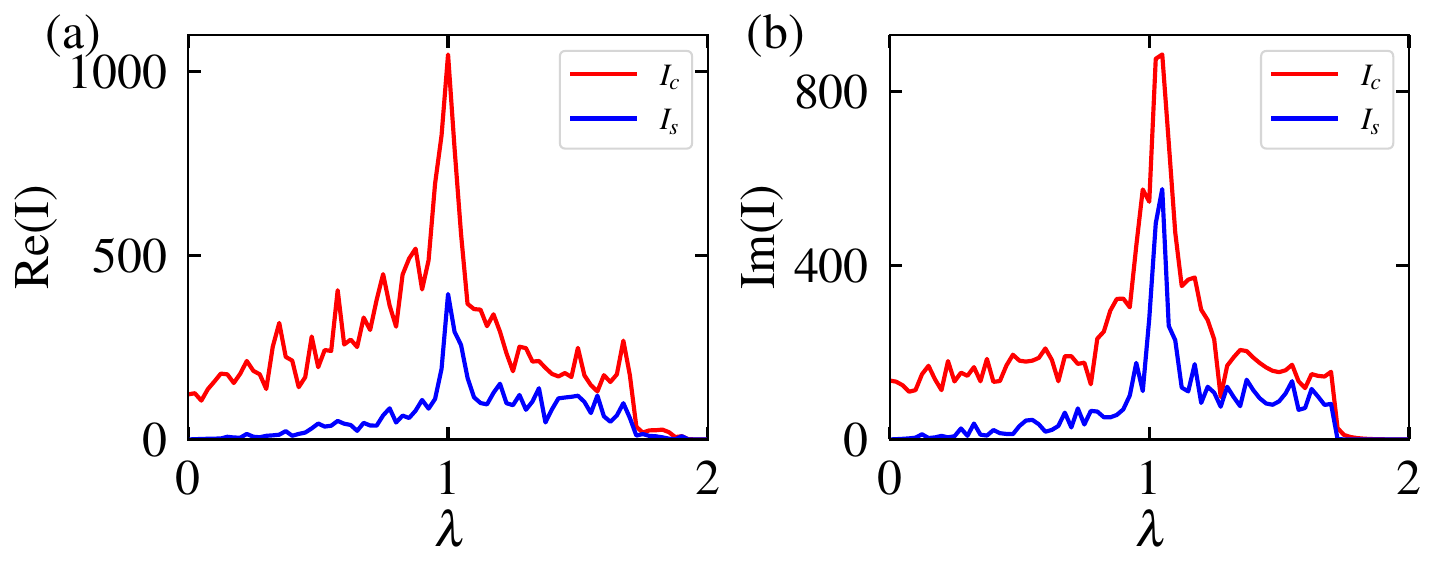}
    \caption{(a) Real and (b) imaginary components of the current versus disorder strength $\lambda$ for $h_z = 0.37$. The red and blue curves correspond to the charge  ($I_c$) and spin current ($I_s$), respectively.
}
    \label{fig:aahlam}
\end{figure}

We now examine the complementary case in which the Zeeman field is fixed and the AAH disorder strength is varied. Figure~\ref{fig:aahlam} shows the real and imaginary components of the current as functions of \( \lambda \) at \( h_z = 0.37 \). For every value of $\lambda$, the current is extracted by taking the maximum magnitude of the flux–dependent current over the entire $2\pi$ flux window, allowing us to capture the strongest response supported by the system. The red and blue curves correspond to the charge and spin currents ($I_c$ and $I_s$), respectively. When AAH modulation is absent ($\lambda=0$), the NH AFM ring exhibits an almost negligible spin current and only a weak charge current, indicating that in the periodic limit, the combination of non-Hermiticity and AFM order alone is insufficient to support sufficient transport. As $\lambda$ is introduced and gradually increased, both $I_c$ and $I_s$ rise in a distinct nonlinear fashion, reflecting how the AAH potential reorganizes the underlying spectrum and opens additional transport pathways. The emergence and subsequent enhancement of the $I_s$ are directly tied to an imbalance between the two spin channels: when the transport in the up (or down) sector begins to dominate significantly over the opposite one, the spin current, which is simply the difference between the up- and down-spin contributions, naturally develops and grows. The quasiperiodic modulation $\lambda$ plays a central role in creating this imbalance, as it modifies the spin-resolved transport in an asymmetric manner and thereby enables a finite and tunable spin-current response. However, once $\lambda$ becomes large enough, strong quasiperiodic scattering suppresses coherent motion, causing both real and imaginary parts of the current to decrease again. This non-monotonic dependence appears in both sectors of the current (real and imaginary).
%, underscoring that the interplay between non-Hermiticity, AFM background, and quasiperiodic modulation crucially controls the transport response. 

\begin{figure}[t]
    \centering
   \includegraphics[width=1.0\linewidth]{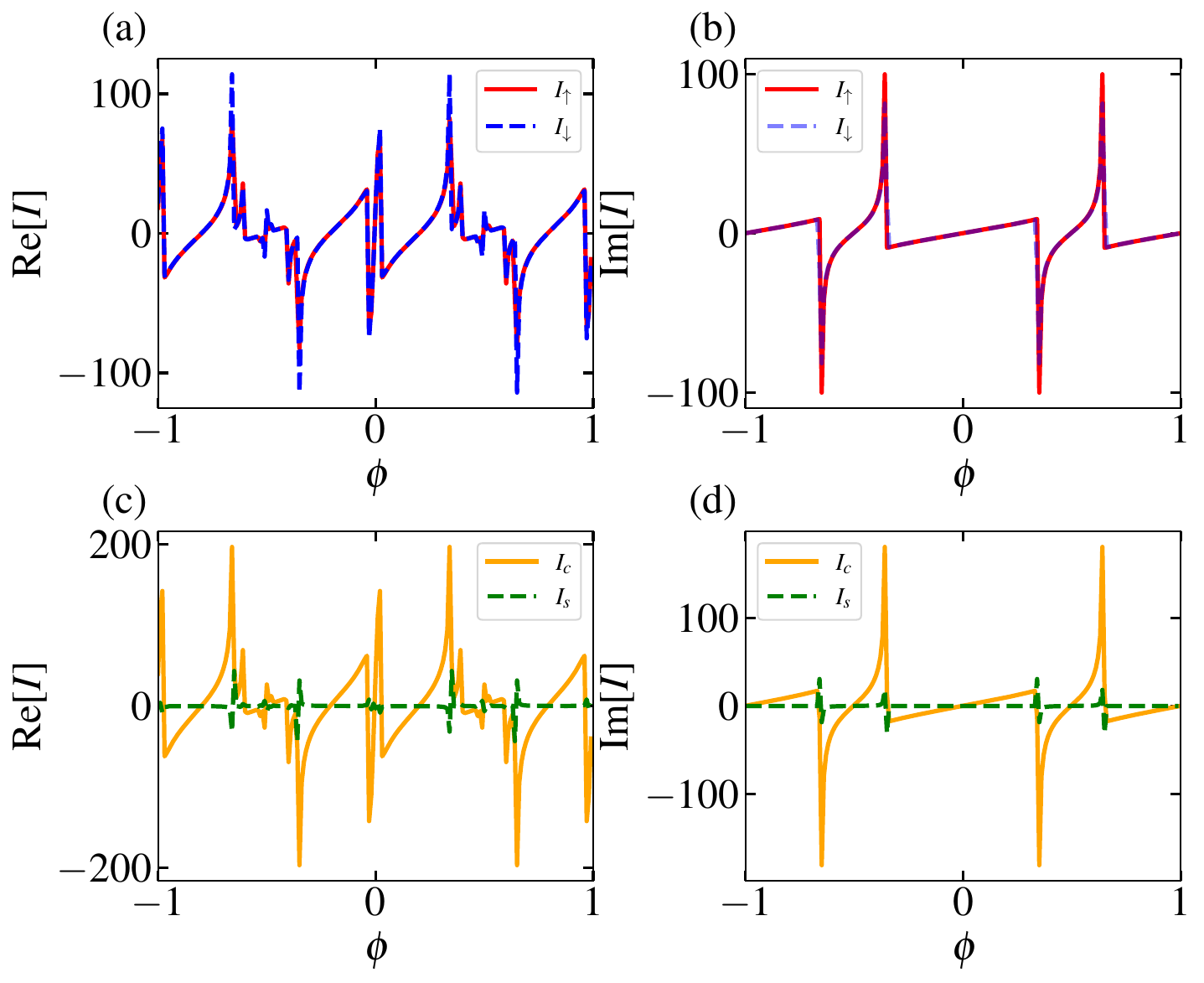}
    \caption{(a) and (b) show the real and imaginary parts of the up and down currents, $I_\uparrow$ (blue dashed line) and $I_\downarrow$ (red solid line), as a function of $\phi$, respectively.  (c) and (d) display the corresponding real and imaginary parts of the charge current $I_c$ (yellow solid line) and spin current $I_s$ (green dashed line) as a function of $\phi$. The strength of Zeeman field and quasiperiodic potential are fixed at $h_z = 0.37$ and $\lambda = 0.5$, respectively.}
    \label{fig:aahcurphi1}
\end{figure}

\begin{figure}[t]
    \centering
   \includegraphics[width=1.0\linewidth]{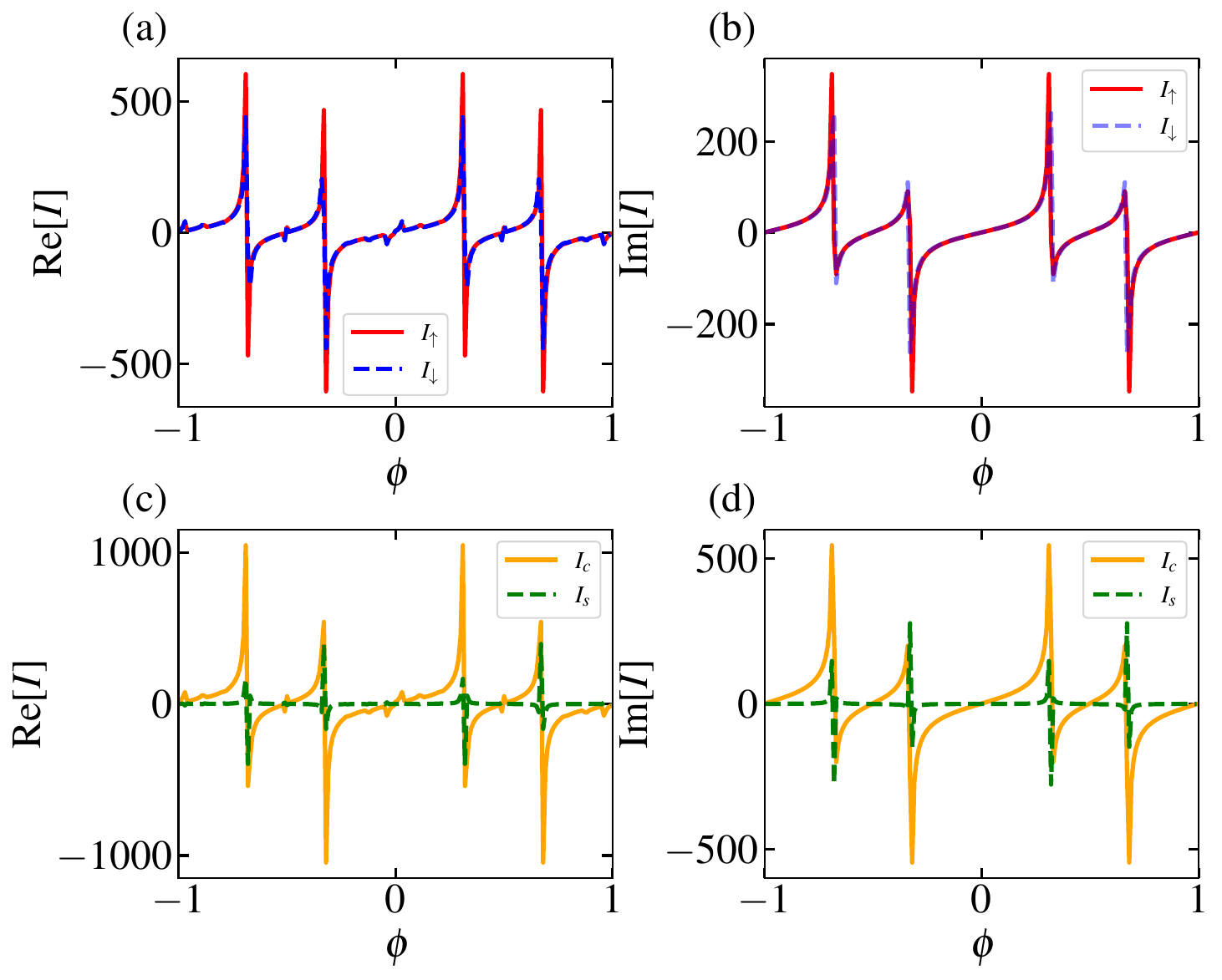}
    \caption{(a) and (b) depict the real and imaginary components of the spin-resolved currents, $I_{\uparrow}$ and $I_{\downarrow}$, respectively, plotted as functions of the magnetic flux $\phi$. The up ($I_{\uparrow}$) and down $I_{\downarrow}$ spin currents are shown by the blue dashed and red solid curves, respectively. (c) and (d) present the corresponding real and imaginary parts of the charge ($I_c$) and the spin ($I_s$) current as functions of $\phi$, with yellow solid and green dashed curves, respectively. Here, $h_z = 0.37$ and $\lambda = 1.0$.}
    \label{fig:aahcurphi2}
\end{figure}

To further clarify the origin of the charge and spin current variations, in Fig.~\ref{fig:aahcurphi1} and Fig.~\ref{fig:aahcurphi2} we resolve the total current into its elementary spin-resolved contributions by plotting the up- and down-spin currents as functions of the auxiliary flux, along with the resulting charge and spin currents constructed from them for different values of $\lambda$. In the figure, the red and blue curves correspond to the up- and down-spin channels, while the orange and green curves represent the charge and spin currents, respectively. When the quasiperiodic strength is set to $\lambda=0.5$, the up- and down-spin currents become nearly indistinguishable over most of the flux interval, leading to an almost vanishing spin current across that entire region. This scenario manifests simultaneously  in both the real and imaginary components of the current, signifying that a moderate degree of quasiperiodicity alone is insufficient to disrupt the balance between the two spin channels. Consequently, the spin-resolved transport remains effectively symmetric, highlighting that quasiperiodic modulation, in the absence of additional symmetry-breaking mechanisms, does not provide the necessary ground to induce appreciable spin polarization in the current. However, upon increasing the disorder strength to $\lambda=1$, the situation changes substantially: the spin-resolved currents begin to diverge appreciably from one another across a wide portion of the flux window. This enhanced asymmetry directly translates into a substantial spin current, as the difference between the up- and down-spin responses becomes pronounced over the dominant flux ranges, giving rise to a robust spin-current signature in both its real and imaginary components.

\begin{figure}[t]
    \centering
    \includegraphics[width=1.0\linewidth]{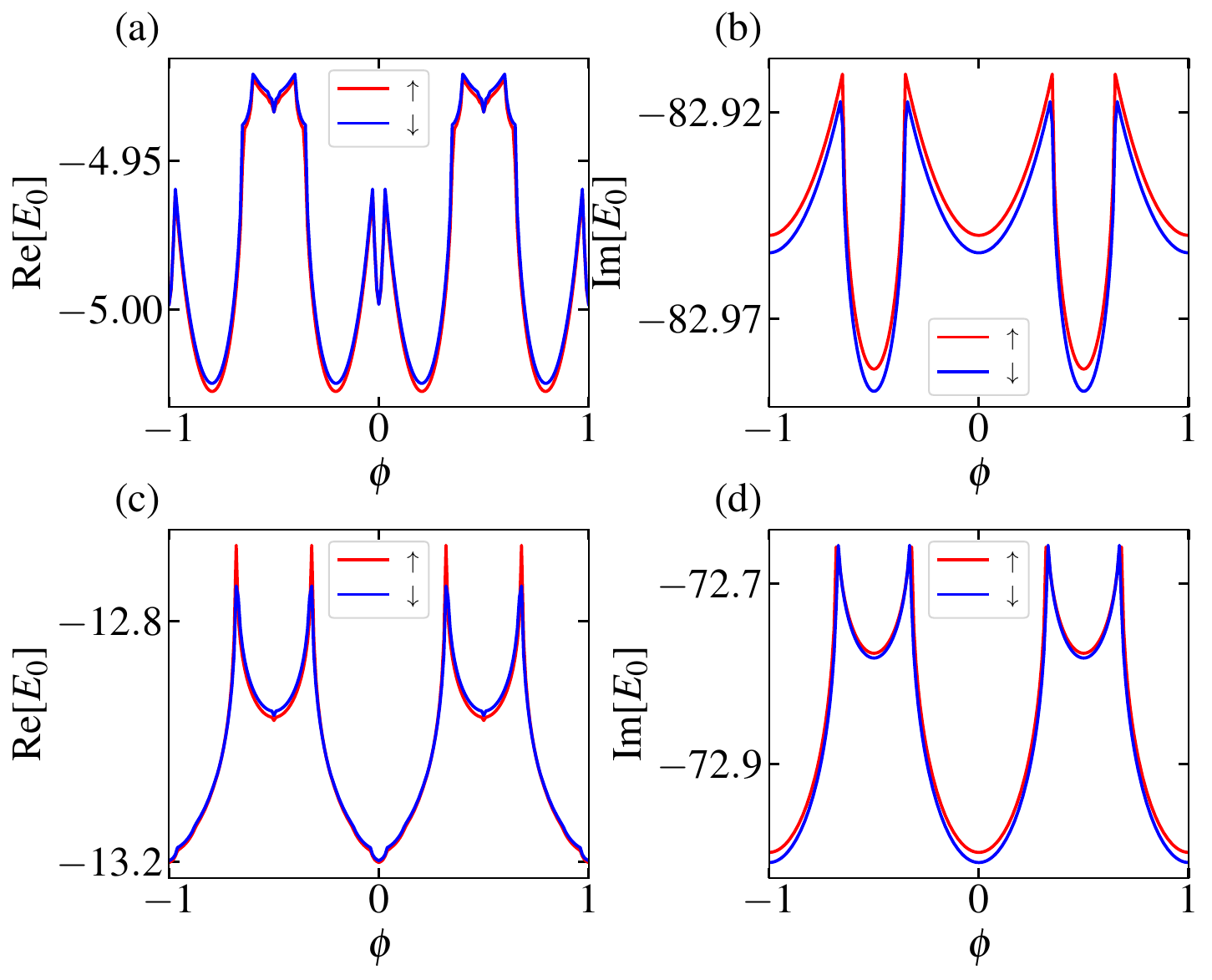}
    \caption{((a,c) Real and (b,d) imaginary components of the ground-state energy versus flux $\phi$. Red and blue solid lines denote the spin-up ($\uparrow$) and spin-down ($\downarrow$) sectors, respectively. Here, $h_z = 0.37$, the upper (lower) panels correspond to $\lambda = 0.5$ ($\lambda = 1.0$).}
    \label{fig:aahground}
\end{figure}

To gain additional insight into the underlying mechanism, as shown in Fig.~\ref{fig:aahground}, we examine the flux dependence of both the real and imaginary components of the ground-state energy for two representative values of the quasiperiodic potential strength $\lambda$, while keeping $h_z$ fixed. This analysis provides direct insight into how the underlying spectral response of the system evolves as the quasiperiodic modulation is strengthened. For the weaker modulation, $\lambda = 0.5$, the real part of the ground-state energy exhibits only a narrow variation with flux, spanning approximately from $-4.925$ to $-5.025$ as shown in Fig.~\ref{fig:aahground}(a). A similar weak flux sensitivity is observed in the imaginary component, which varies in a small window between $-82.98$ and $-82.92$ [see Fig.~\ref{fig:aahground}(b)]. The limited spread in both components indicates that the energy levels respond rather weakly to the applied flux, signaling a relatively suppressed current response in this regime. In contrast, when the quasiperiodic strength is increased to $\lambda = 1.0$, the flux-induced modulation of the spectrum becomes significantly more pronounced. The real part of the ground-state energy now varies over a much wider range, extending from roughly $-12.8$ to $-13.2$, while the imaginary part changes from about $-73.0$ to $-72.7$ [see Fig.~\ref{fig:aahground}(c) and (d)]. This enhanced spectral span reflects a stronger sensitivity of the ground state to the magnetic flux. From a physical viewpoint, the enhanced energy modulation with flux at larger \( \lambda \) gives rise to an increased slope in the energy-flux characteristics. Since the current is proportional to the derivative of the ground-state energy with respect to flux, the larger slope observed at $\lambda = 1.0$ implies a substantially higher current compared to the $\lambda = 0.5$ case. Thus, strengthening the quasiperiodic potential amplifies the flux responsiveness of both the real and imaginary energy spectra, leading to an enhanced current-carrying capability of the system.

\begin{figure}[t]
    \centering
    \includegraphics[width=1.0\linewidth]{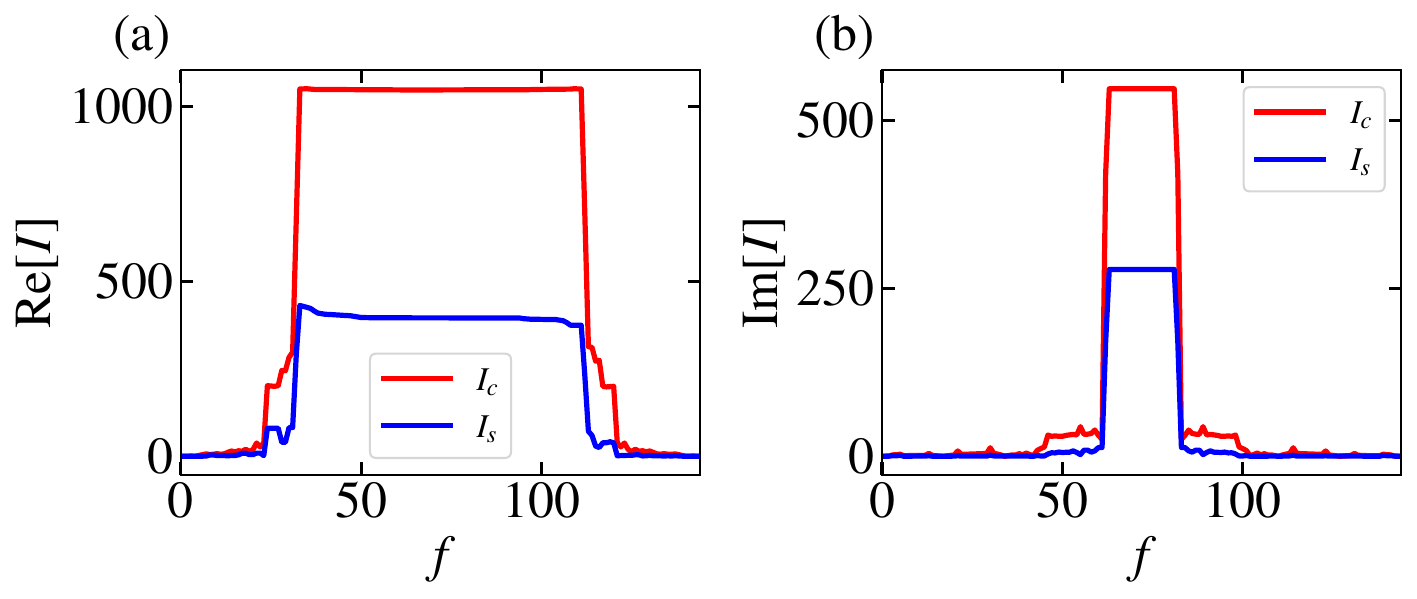}
    \caption{The real and imaginary components of the current are shown as functions of the filling fraction $f$ in (a) and (b), respectively. The charge ($I_c$) and spin ($I_s$) current are represented by the blue and red solid curves, respectively. The values of the parameters are chosen as $h_z = 0.37$ and $\lambda = 1.0$.}
    \label{fig:aahfillingfraction}
\end{figure}

After clarifying the role of the quasiperiodic potential and Zeeman field in shaping the flux response of the current for half filling, we next examine how the current depends on filling. In Fig.~\ref{fig:aahfillingfraction}, we present the filling dependence of both the real and imaginary components of the current, where the red and blue curves correspond to the charge and spin currents, respectively. A salient feature emerging from this non-Hermitian setting, rarely encountered in conventional Hermitian systems, is the appearance of extended filling intervals over which the current remains nearly constant. Such plateau-like behavior is observed in both the real and imaginary sectors, although the extent of these windows differs markedly. The real current exhibits a substantially broader plateau, maintaining an almost maximal value over a wide filling range extending approximately from quarter filling to about three-quarter filling. In contrast, the imaginary current displays a much narrower window, reaching its peak predominantly around half filling. Physically, this distinction reflects the different ways in which real and imaginary spectral components respond to changes in particle occupation: while the real current is governed by the cumulative contribution of many occupied states and hence remains robust over a wide filling interval, the imaginary current is more sensitive to the detailed redistribution of states near the band center, where non-Hermitian effects such as gain–loss-induced asymmetry are most pronounced.
%Consequently, the filling-driven evolution of current reveals a clear separation between the stability of transport encoded in the real part and the more localized, filling-sensitive behavior associated with the imaginary component.

\begin{figure}[t]
    \centering
    \includegraphics[width=1.0\linewidth]{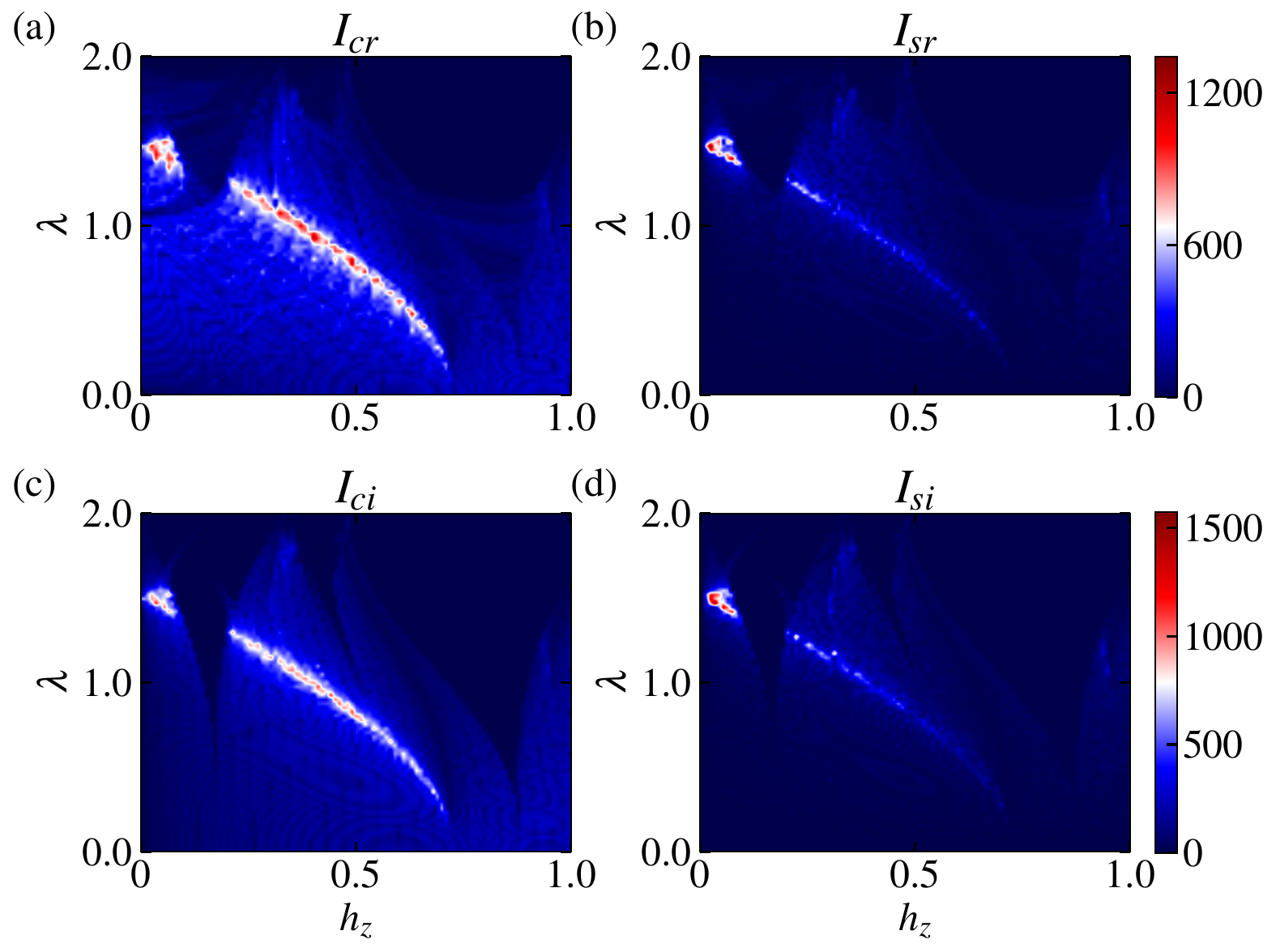}
    \caption{Real and imaginary parts of the charge and spin currents as functions of $h_z$ and $\lambda$. (a) Real part of the charge current $I_{cr}$, (b) real part of the spin current $I_{sr}$, (c) imaginary part of the charge current $I_{ci}$, and (d) imaginary part of the spin current $I_{si}$.} 
    \label{fig:aahphase}
\end{figure}

To obtain a comprehensive picture of the combined influence of $\lambda$ and $h_z$ on transport, we present the complete parameter dependence of the charge and spin currents in Fig.~\ref{fig:aahphase}. In these maps, $h_z$ is varied along the horizontal axis and $\lambda$ along the vertical axis, while the magnitude of the current is encoded through a color scale, with red (blue) representing the maximum (minimum) values and intermediate colors smoothly interpolating between high and low current regions. Several notable features emerge from this two-parameter landscape. First, both charge and spin currents can reach substantially large magnitudes, approaching values of order $1200$ in the real sector and nearly $1500$ in the imaginary sector, underscoring the strong flux sensitivity induced by the combined action of quasiperiodicity and non-Hermiticity. Furthermore, for fixed $\lambda$ cuts, the real current exhibits pronounced re-entrant behavior in the higher-$\lambda$ regime, where transport repeatedly switches between finite and suppressed values as $h_z$ is varied; in contrast, the imaginary current displays similar re-entrant features already at comparatively lower $\lambda$. Another striking observation is that in the regime of small $h_z$ and large $\lambda$, the spin current becomes comparable to the charge current in the real sector and can even exceed it in the imaginary sector, indicating that non-Hermitian effects can strongly amplify spin-selective transport and, in certain parameter windows, invert the conventional hierarchy between charge and spin responses.

\section{Conclusion}
\label{sec:conc}

In conclusion, we have systematically investigated charge and spin transport in a non-Hermitian spinful quantum ring in the presence of structured Zeeman fields and quasiperiodic modulations. We demonstrated that non-Hermiticity, implemented via anti-Hermitian hopping, generates an intrinsic synthetic flux that fundamentally reshapes the energy spectrum and transport response. While ferromagnetic and nonmagnetic configurations support only imaginary circulating currents with vanishing charge and spin responses, an antiferromagnetic Zeeman texture gives rise to finite real and imaginary charge currents, highlighting the crucial role of staggered spin ordering in enabling transport. In the clean limit the exact cancellation between spin-up and spin-down contributions suppresses spin transport despite the presence of non-Hermiticity.

Upon introducing quasiperiodic onsite potential modulation, this spin symmetry is lifted, leading to the simultaneous emergence and strong enhancement of spin and charge currents under balanced spin populations. We showed that quasiperiodicity reorganizes the complex energy spectrum and enhances its sensitivity to the effective flux, resulting in large, tunable transport responses and re-entrant current behavior as functions of the Zeeman field and disorder strength. 
Looking ahead, several promising directions naturally follow from this work. It would be highly interesting to explore the robustness of the observed transport phenomena against electron–electron interactions and finite-temperature effects, as well as to examine the role of non-Hermitian topology in shaping spin-selective currents.

\section{Acknowledgment}
T.M. acknowledges support from Science and Engineering Research Board (SERB), Govt. of India, through project No. MTR/2022/000382 and STR/2022/000023.

\appendix

\section{Dispersion relation for nonmagnetic and ferromagnetic case}

One can diagonalize the Bloch Hamiltonian described in Eq. (\ref{eq:FM_Bloch_fg}) to get the dispersion relation and the associated current. Since hopping conserves spin, the Hamiltonian decouples into two independent
$2\times2$ blocks corresponding to the spin sectors
$(A\uparrow,B\uparrow)$ and $(A\downarrow,B\downarrow)$. Each block can be
written as
\begin{equation}
\mathcal{H}_s(k)
=
\begin{pmatrix}
 s h_z & f(k) \\
 g(k) & s h_z
\end{pmatrix},
\qquad s=\pm1 .
\end{equation}

The eigenvalues follow from the characteristic equation
\begin{equation}
\det\!\left[\mathcal{H}_s(k)-E\right]
=
(E-s h_z)^2 - f(k) g(k)=0 ,
\end{equation}
yielding
\begin{equation}
E_{s,\pm}(k)
=
s h_z
\pm
\sqrt{f(k) g(k)} .
\end{equation}

For the above parametrization, one finds explicitly
\begin{equation}
\begin{aligned}
f(k) g(k)
&=
\left(t e^{i\phi^{\prime}}-t e^{-i(\phi^{\prime}+k)}\right)
\left(-t e^{-i\phi^{\prime}}+t e^{i(\phi^{\prime}+k)}\right) \\
&=
-4 t^2 \sin^2\!\left(\frac{k+2\phi^{\prime}}{2}\right) ,
\end{aligned}
\end{equation}
which is negative definite. Consequently,
\begin{equation}
\sqrt{f(k) g(k)}
=
2 i t \sin\!\left(\frac{k+2\phi^{\prime}}{2}\right) .
\end{equation}

The full dispersion relation in terms of the Bloch momentum reads
\begin{equation}
\boxed{
E_{s,\pm}(k)
=
s\,h_z
\pm
2 i t
\sin\!\left(\frac{k+2\phi^{\prime}}{2}\right),
  s=\pm1.
}
\end{equation}

Thus, in the ferromagnetic case the uniform Zeeman field rigidly shifts the
spin-resolved bands by $\pm h_z$, while the asymmetric non-Hermitian hopping
drives the spectrum off the real axis and generates purely imaginary
dispersive branches. The four bands originate from the combined spin and
artificial sublattice degrees of freedom, without any Brillouin-zone folding. Since each unit cell contains two lattice sites, the energy bands are twofold degenerate.

In the absence of the Zeeman field, $h_z=0$, spin degeneracy is restored and the dispersion reduces to
\begin{equation}
E_{\pm}(k)
=
\pm
2 i t
\sin\!\left(\frac{k}{2}+\frac{2\pi\phi}{L}\right),
\end{equation}
where each branch is twofold degenerate in spin.

The current is obtained by differentiating the energy with respect to the flux $\phi$,
\begin{equation}
I_{s,\pm}(k)
=
-\,c\frac{\partial E_{s,\pm}(k)}{\partial \phi} .
\end{equation}
Using
\begin{equation}
E_{s,\pm}(k)
=
s\,h_z
\pm
2 i t
\sin\!\left(\frac{k}{2}+\frac{2\pi\phi}{L}\right),
\end{equation}
we obtain
\begin{equation}
\frac{\partial E_{s,\pm}}{\partial \phi}
=
\pm\,\frac{4\pi}{L} i t
\cos\!\left(\frac{k+2\phi}{2}\right),
\end{equation}
which yields the current
\begin{equation}
\boxed{
I_{s,\pm}(k)
=
\mp\,\frac{4\pi c}{L} i t
\cos\!\left(\frac{k}{2}+\frac{2\pi\phi}{L}\right).
}
\end{equation}

% #-------------------

\section{Dispersion relation for anti-ferromagnetic case}
We now consider the antiferromagnetic configuration, where the Zeeman
field alternates in sign between consecutive sites. For spin $\uparrow$ and
$\downarrow$, the on-site energies on the two sublattices $(A,B)$ are
\begin{align*}
\text{A: } & +h_z,\;-h_z, \\
\text{B: } & -h_z,\;+h_z ,
\end{align*}
which results in a genuine two-site unit cell. Thus, the Bloch Hamiltonian is described in Eq. (\ref{eq:AFM_Bloch_fg}).
Since hopping conserves spin, the Hamiltonian decouples into two identical
$2\times2$ blocks for spin $\uparrow$ and $\downarrow$. Each block takes the
form
\begin{equation}
\mathcal{H}_{\mathrm{AFM}}^{(s)}(k)
=
\begin{pmatrix}
 s h_z & f(k) \\
 g(k) & -s h_z
\end{pmatrix},
\qquad s=\pm1 .
\end{equation}

The eigenvalues are obtained from the characteristic equation
\begin{equation}
\det\!\left[\mathcal{H}_{\mathrm{AFM}}^{(s)}(k)-E\right]
=
E^2 - h_z^2 - f(k) g(k)=0 ,
\end{equation}
leading to
\begin{equation}
E_{s,\pm}(k)
=
\pm \sqrt{\,h_z^2 + f(k) g(k)} .
\end{equation}

Substituting the explicit forms of $f(k)$ and $g(k)$, one finds
\begin{equation}
\begin{aligned}
f(k) g(k)
&=
\left(t e^{i\phi^{\prime}}-t e^{-i(\phi^{\prime}+k)}\right)
\left(-t e^{-i\phi^{\prime}}+t e^{i(\phi^{\prime}+k)}\right) \\
&=
-4 t^2 \sin^2\!\left(\frac{k+2\phi^{\prime}}{2}\right) .
\end{aligned}
\end{equation}
The dispersion relation therefore becomes
\begin{equation}
\boxed{
E_{s,\pm}(k)
=
\pm
\sqrt{
h_z^2
-
4 t^2
\sin^2\!\left(\frac{k}{2}+\phi^{\prime}\right)
},
\qquad s=\pm1 .
}
\end{equation}

The four bands arise from the combined spin and true sublattice degrees of
freedom. In contrast to the ferromagnetic case, the staggered Zeeman field
enters quadratically in the spectrum and opens a gap at the Brillouin-zone
boundary. As a result, although the hopping is non-Hermitian, the AFM
configuration supports real quasienergy bands as long as
$h_z^2 > 4 t^2 \sin^2[(k+2\phi^{\prime})/2]$,
\begin{equation}
\begin{cases}
E_{s,\pm}(k) = \pm \sqrt{\Delta(k+\phi^{\prime})}, 
& |h_z|\ge 2|t|\left|\sin\!\left(\frac{k}{2}+\phi^{\prime}\right)\right|,\\[4pt]
E_{s,\pm}(k) = \pm i \sqrt{-\Delta(k+\phi^{\prime})},
& |h_z|< 2|t|\left|\sin\!\left(\frac{k}{2}+\phi^{\prime}\right)\right|,
\end{cases}
\end{equation}
where, $\Delta(k+\phi^{\prime})=h_z^2 - 4t^2\sin^2(\frac{k}{2} + \phi^\prime)$.

Using
\begin{equation}
E_{\pm}(k)
=
\pm\, i
\sqrt{
4 t^2
\sin^2\!\left(\frac{k}{2}+\frac{2\pi\phi}{L}\right)
-
h_z^2
}.
\end{equation}

Differentiating with respect to the flux $\phi^{\prime}$, we obtain
\begin{equation}
\frac{\partial E_{\pm}}{\partial \phi}
=
\pm\, i\,
\frac{
\frac{2\pi}{L}4 t^2
\sin\!\left(\frac{k}{2}+\frac{2\pi\phi}{L}\right)
\cos\!\left(\frac{k}{2}+\frac{2\pi\phi}{L}\right)
}{
\sqrt{
4 t^2
\sin^2\!\left(\frac{k}{2}+\frac{2\pi\phi}{L}\right)
-
h_z^2
}
}.
\end{equation}

The corresponding current,
$I_{\pm}(k)=-\partial E_{\pm}(k)/\partial \phi$,
is therefore given by
\begin{equation}
\boxed{
I_{\pm}(k)
=
\mp\, ic\,
\frac{
\frac{4\pi}{L} t^2
\sin\!\left(k+\frac{4\pi\phi}{L}\right)
}{
\sqrt{
4 t^2
\sin^2\!\left(\frac{k}{2}+\frac{2\pi\phi}{L}\right)
-
h_z^2
}
}.
}
\end{equation}
Now for
\begin{equation}
E_{s,\pm}(k)
=
\pm
\sqrt{
h_z^2
-
4 t^2
\sin^2\!\left(\frac{k}{2}+\frac{2\pi\phi}{L}\right)
},
\qquad s=\pm1 .
\end{equation}

The corresponding current,
$I_{\pm}(k)=-\partial E_{s,\pm}(k)/\partial\phi$,
is given by
\begin{equation}
\boxed{
I_{\pm}(k)
=
\pm c
\frac{
\frac{4\pi}{L} t^2
\sin\!\left(k+\frac{4\pi\phi}{L}\right)
}{
\sqrt{
h_z^2
-
4 t^2
\sin^2\!\left(\frac{k}{2}+\frac{2\pi\phi}{L}\right)
}
}.
}
\end{equation}

\bibliography{ref}

\end{document}